\documentclass[usenatbib]{mn2e}

\usepackage{epsfig}
\usepackage{color}
\usepackage{caption}
\usepackage{aas_macros}  
\usepackage{amsmath,amssymb}
\usepackage{chngcntr}

\newcommand{\Msolar}{\mbox{\,$\rm M_{\odot}$}}        
\newcommand{\Lsolar}{\mbox{\,$\rm L_{\odot}$}}        

\newcommand{\Jstar}{\mbox{\,J1845$-$4138}}

  \newcommand{\Teff}{\mbox{\,\em T$_{\rm eff}$}}         
 \newcommand{\teff}{\mbox{\,$T_{\rm eff}$}}      
\newcommand{\lgcs}{\mbox{\,$\log g / {\rm cm\,s^{-2}}$}}        
\newcommand{\yy}{\mbox{\,$n_{\rm He}/n_{\rm H}$}}        

  \newcommand{\kmsec}{\,\mbox{$\mbox{km}\,\mbox{s}^{-1}$}}    
  \def\simge{\mathrel{\raise1.16pt\hbox{$>$}\kern-7.0pt
    \lower3.06pt\hbox{{$\scriptstyle \sim$}}}}           
  \def\simle{\mathrel{\raise1.16pt\hbox{$<$}\kern-7.0pt
    \lower3.06pt\hbox{{$\scriptstyle \sim$}}}}           
\title[GALEX J1845$-$4138: a new extreme helium star]{GALEX J184559.8$-$413827: a new extreme helium star identified using SALT\thanks{based on an observation made with the Southern African Large Telescope (SALT)} }
\author[C. S.~Jeffery]{C. Simon Jeffery$^{1,2}\thanks{email: csj@arm.ac.uk}$\\
$^{1}$Armagh Observatory and Planetarium, College Hill, Armagh BT61 9DG, UK\\
$^{2}$School of Physics, Trinity College Dublin, College Green, Dublin 2, Ireland\\
}
\begin{document}

\date{Accepted \ldots. Received \ldots; in original form \ldots}

\pagerange{\pageref{firstpage}--\pageref{lastpage}} \pubyear{2014}

\maketitle

\label{firstpage}

\begin{abstract}
A high-resolution spectrum of the helium-rich `hot subdwarf' GALEX J184559.8$-$413827 (\Jstar) obtained with SALT HRS demonstrates it to be the first extreme helium (EHe) star to be discovered in nearly 40 years. 
 A quantitative analysis demonstrates it to have an atmosphere described by $T_{\rm eff} = 26\,170 \pm750$\,K,  $\log g /{\rm cm\,s^{-2}} = 4.22\pm 0.10$, and a surface chemistry characterised by CNO-processed helium, a 1\% contamination of hydrogen  (by number), and a metallicity 0.4 dex subsolar. 
Its distance and position are consistent with membership of the Galactic bulge. 
Its sharp absorption lines place strong constraints on both the rotation {\it and} microturbulent velocities.     
Spectroscopically, \Jstar\ closely resembles the pulsating EHe star V652\,Her, generally considered to be the product of a double helium white dwarf merger evolving to become a helium-rich sdO star. 
\end{abstract}

\begin{keywords}
             stars: abundances,
             stars: fundamental parameters,
             stars: chemically peculiar,
             stars: individual (GALEX\,J184559.8$-$413827, V652\,Her)
             \end{keywords}

\section{Introduction}
\label{s:intro} Extreme helium stars (EHes) comprise some seventeen stars of spectral types equivalent to 
A and B but having weak or no hydrogen Balmer lines. In their place, relatively sharp and 
strong lines of neutral helium indicate low surface gravities and atmospheres dominated by helium. The next most abundant elements are carbon, nitrogen and oxygen, indicative of highly processed material  exposed at the stellar surface. 
The first extreme helium star, HD124448, was identified by \citet{popper42}. 
This was followed by discoveries spanning some thirty years, many arising from  spectroscopic surveys of luminous blue stars. 
All of these have been analysed spectroscopically in more 
or less detail, giving effective temperature and surface gravities; a full list  is provided by \citet{jeffery08.ibvs}, to which should be added analyses of BD+10$^{\circ}$2179 by \citet{pandey11} and \citet{kupfer17}.  
Translating surface gravity into luminosity-to-mass ratio ($L/M$), the class covers a range of nearly 2 dex in  $L/M$, with the brightest members lying close to the Eddington limit. 
Hence, the fact that there have been no new luminous EHes discovered since about 1980 \citep{drilling86} is easily understood. 
With surveys of luminous blue stars  complete to $B=12$ for the Milky Way and for $b=\pm30^{\circ}$ and $l=\pm60^{\circ}$ \citep{drilling95,drilling96}, fainter EHes would either lie well outside the Milky Way or be heavily obscured and reddened by its central bulge. 
At the least luminous end of their range, EHe stars have surface gravities similar to those seen on the main-sequence. 
The EHes appear to be substantially less numerous. 

Since all of the classical EHes were discovered, there have been numerous spectroscopic surveys for {\it faint} blue stars (sometimes masquerading as quasar or galaxy surveys), including the Palomar-Green \citep{green86},  Hamburg quasar and Hamburg/ESO  \citep{hagen95,wisotzki96},  Edinburgh-Cape \citep{stobie97a}, HK \citep{beers92},   Sloan  \citep{york00} and GALEX \citep{bianchi14} surveys.  Each adopted a scheme to classify the spectra of stellar objects and thereby identified many new and interesting stars. 
 \citet{drilling96} recognized that the  class identified by \citet{green86} as  sdOD (``pure He {\sc i} absorption spectra, characterized by the weakness or absence of hydrogen Balmer lines and He {\sc ii} 4686 while showing the singlet He {\sc i} 4388 about equal in strength to the triplet He {\sc i} 4471''), and by \citet{moehler90} as He-sdB, is the same definition as that given for extreme helium stars and binaries by \citet{drilling86}. Yet no new EHe stars were found. 


The sdOD/He-sdB classification also embraces helium-rich stars of higher surface gravity, i.e. helium-rich hot subdwarfs. 
Considerable effort has been spent on discovering and exploring these stars, with  spectacular results.
They turn out to fall into diverse groups, including the double subdwarf PG1544+488 \citep{schulz91,ahmad04b,sener14},  
the binary CPD-20$^{\circ}1123$ \citep{naslim12}, 
through the chemically-peculiar intermediate helium subdwarfs \citep{naslim11,naslim13,jeffery17a},  
the low-gravity He-sdO stars and O(He) stars \citep{husfeld89,reindl14}
to the high-gravity extreme helium sdO stars 
\citep{greenstein74,dreizler90,stroeer07,naslim10,justham11,zhang12a}.   
\citet{drilling96} and \citet{drilling13} introduced an MK-like classification scheme  which clearly distinguishes between helium-rich subdwarfs and extreme helium stars, but requires a resolution closer to  1.5\AA\ than the 10\AA\ of many surveys. 
Hence, survey spectra have rarely been classified using  the Drilling scheme\footnote{A recent exception used tools in the Virtual Observatory \citep{perez16}.}.
Stars of interest to us have more frequently been classified as He-sdB, He-sdOB, or He-sdO

Efforts to find more examples of some subclasses require high-resolution and high signal-to-noise spectra to identify weak  and double lines. 
A  spectroscopic survey of stars classified as helium-rich subdwarfs,  especially He-sdB or sdOD using 8m class telescopes has been undertaken.
This paper reports an observation of one such star obtained with the high-resolution spectrograph (HRS)
on the  Southern Africa Large Telescope (SALT) \citep{buckley06,bramall10,crause14}


\begin{figure*}
       \epsfig{file=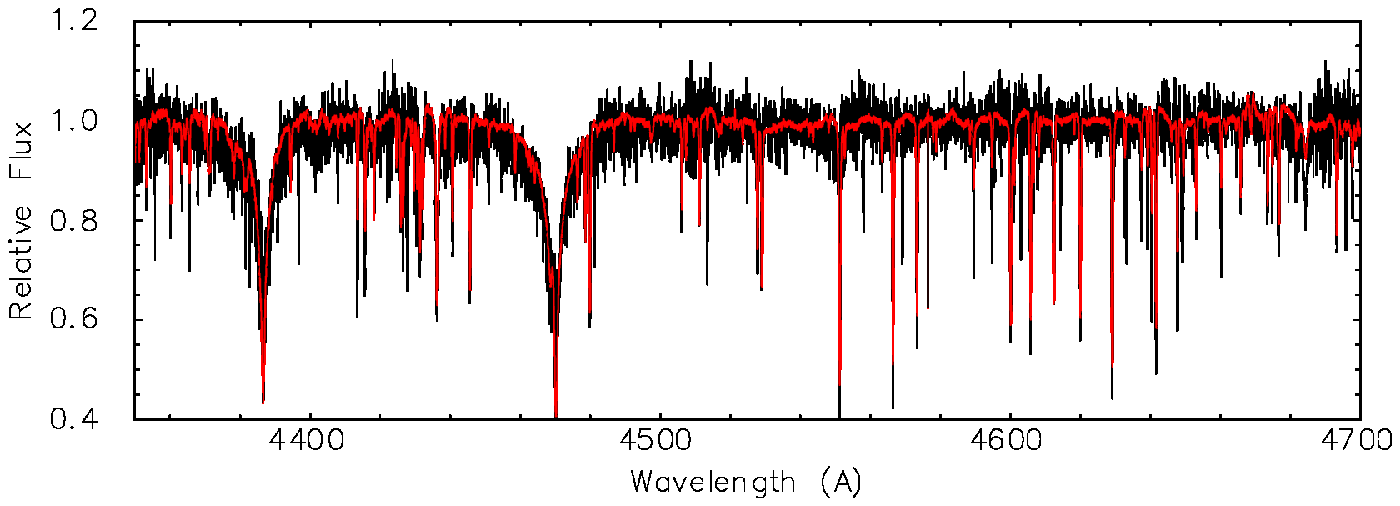,width=170mm,clip=}
        \caption{Part of the renormalised SALT HRS spectrum of \Jstar\ (black) compared with the median spectrum of V652\,Her near maximum radius obtained by \citet[red]{jeffery15b}. } 
        \label{f:lores}
\end{figure*}

\section[]{Observations}
\label{s:obs}

GALEX J184559.8$-$413827 ($\alpha_{2000} = 18^{\circ} 45\arcmin 59.8\arcsec$, $\delta_{2000} = 41^{\circ} 38\arcmin 27\arcsec$, $V=14.6$:   \Jstar\ hereafter ) was identified to be a faint blue star from GALEX colours by \citet{vennes11}.  A flux-calibrated low-resolution spectrum obtained with the Faint Object Spectrograph and Camera mounted on the European Southern Observatory's New Technology Telescope (EFOSC2/NTT)  on 2008 October 21 was classified `He-sdB', being dominated by neutral helium lines. 
 A coarse analysis of the spectrum and colours yielded first an effective temperature  $T_{\rm eff}=36\,400\pm3\,200$\,K, surface gravity $\log g /{\rm cm\,s^{-1}}=5.75\pm0.65$, and surface helium-to-hydrogen ratio $>0.40$ \citep{vennes11}, and second $T_{\rm eff}=35\,930^{+840}_{-4770}$\,K,  $\log g /{\rm cm\,s^{-1}}=5.75^{+0.27}_{-0.23}$,  $\log \yy=2.10^{+1.10}_{-0.38}$ \citep{nemeth12}.  Both studies used the same observations; the first used  atmospheres comprising hydrogen and helium only, but in which departures from local thermodynamic equilibrium were considered (non-LTE) , while the second used non-LTE models comprising hydrogen, helium, carbon, nitrogen and oxygen. The latter also provided weak upper limits for the CNO abundances relative to hydrogen of 15\%, 75\% and 22\% respectively. 

The `He-sdB' classification led to the inclusion of \Jstar\ on a target list for observations of chemically-peculiar hot subdwarfs with the SALT HRS. One observation was obtained at UTC 02:24 on 2017 March 17  with a median seeing of 1.3\arcsec.  Two exposures were obtained in each of the HRS cameras, the first with an exposure time of  1300\,s, the second was terminated after 700\,s by a technical problem. A second observation was attempted the following night, but the wrong star was observed. 

The HRS spectrum was reduced  to 1-dimension (total counts versus wavelength)  object and sky orders using the SALT HRS pipeline\footnote{PyHRS: http://pysalt.salt.ac.za/} \citep{crawford16}. 
Flux calibration was not attempted.  
The wavelength ranges covered by the spectra are 3860 -- 5519 \AA\    and 5686 -- 8711 \AA\ respectively, the combined  spectrum having a S/N ratio in the range 20 -- 30 at a resolution of $\approx 37,000$. 
The orders were rectified, mapped onto a common wavelength grid (equally spaced in log wavelength)  and merged using an order-management tool written for \'echelle spectra by the author.   
Rectification at low exposure levels remains problematic, in the current case leading to problems at the  blue end of the blue spectrum.

Fig.\,\ref{f:lores} shows that the He\,{\sc i} absorption lines are readily apparent and everywhere stronger  than the Balmer lines. He\,{\sc ii} 4686\AA, can be  identified,  but not the He\,{\sc ii}  Pickering series. 
 The  He\,{\sc i} lines are everywhere narrower than seen in helium-rich hot subdwarfs with 
 $\log g \geq 5$, but less sharp than the lowest-gravity extreme helium stars. 
 On the other hand the spectrum is remarkably {\it similar} to that of  the  better known EHe star V652\,Her \citep{jeffery15b}, including in particular the extremely rich spectrum of singly-ionized nitrogen (Fig.\,\ref{f:lores}).
 The most marked difference is that He\,{\sc ii} 4686\AA\ is significantly stronger in \Jstar, indicating a higher    $T_{\rm eff}$. A  detailed analysis to determine the surface properties more precisely is presented in the next section. 

A single value for the heliocentric radial velocity of \Jstar\ was measured by comparing the observed spectrum  obtained at 2017 03 17 02h 24m (Julian Date 2457828.6000) with the best-fit theoretical spectrum, yielding a value  $-66.1\pm0.2$\kmsec.  There was no discernible shift between the two exposures.  The mean value is approximately consistent with a series of five measurements made over the space of three nights in 2012 and which gave $\bar{v}=-57.6\pm 6.1 \kmsec$ \citep{kawka15}. 

%
%
\begin{table*}
\caption[Atmospheric abundances]
   {Atmospheric abundances of  {\Jstar}, helium stars with similar $L/M$ ratios, and the Sun.
    Abundances are given as $\log \epsilon$, normalised to $\log \Sigma \mu \epsilon = 12.15$. }
\label{t:abunds}
\small
\begin{center}
\begin{tabular}{lrrrrrrrrrrrrrrr}
Star            & $\log \epsilon$    &     &    &    &    &    &    &    &    &    &    &    &  
  &   Ref  \\
      &    H &  He &  C &  N &  O & Ne & Mg & Al & Si &  P &  S &  A & 
Fe &  \\[1mm]
{\Jstar} & 9.56 & 11.54 & 6.91 & 8.69 & 7.78 & 8.29 & 7.91 & 6.44 & 7.61 & 5.62 & 6.93 & -- & 7.07 & \\ 
{V652\,Her}     &9.61 &11.54&7.29 &8.69 &7.58&7.95 &7.80&6.12&7.47&6.42:&7.05&6.64&7.04& 1,2  \\[1mm]
{BX\,Cir}       & 8.1 &11.5 &9.02 &8.4  &8.0 &    &    &7.2 &6.0 &6.8 &5.0 & 6.6& 6.6 & 3 \\
{LS\,IV$+6^{\circ}2$} & 7.3 & 11.52 & 9.41 & 8.54 & 8.30 & 9.35 & 7.34 & 6.26 & 7.11 & 5.99 & 6.99 & & 7.10 & 4 \\
{HD144941}   &10.3 &11.5 &6.80 &6.5  &7.0 &    &    &6.1 &4.8 &6.0 &    &    & 5.7 & 5 \\
{Sun}    &12.0 &10.93 &8.43 &7.83 &8.69&7.93&7.60&6.45&7.51&5.45&7.12&6.40&7.50 & 6 \\[2mm]
\end{tabular}
\end{center}
\parbox{170mm}{
{: value uncertain. } \\
References. 
1~\citet{jeffery99},
2:~\citet{jeffery01b},
3:~\cite{drilling98}, 
4:~\cite{jeffery98},
5:~\citet{harrison97,jeffery97},
6:~\citet{asplund09}.
}
\end{table*}

\begin{figure}
       \epsfig{file=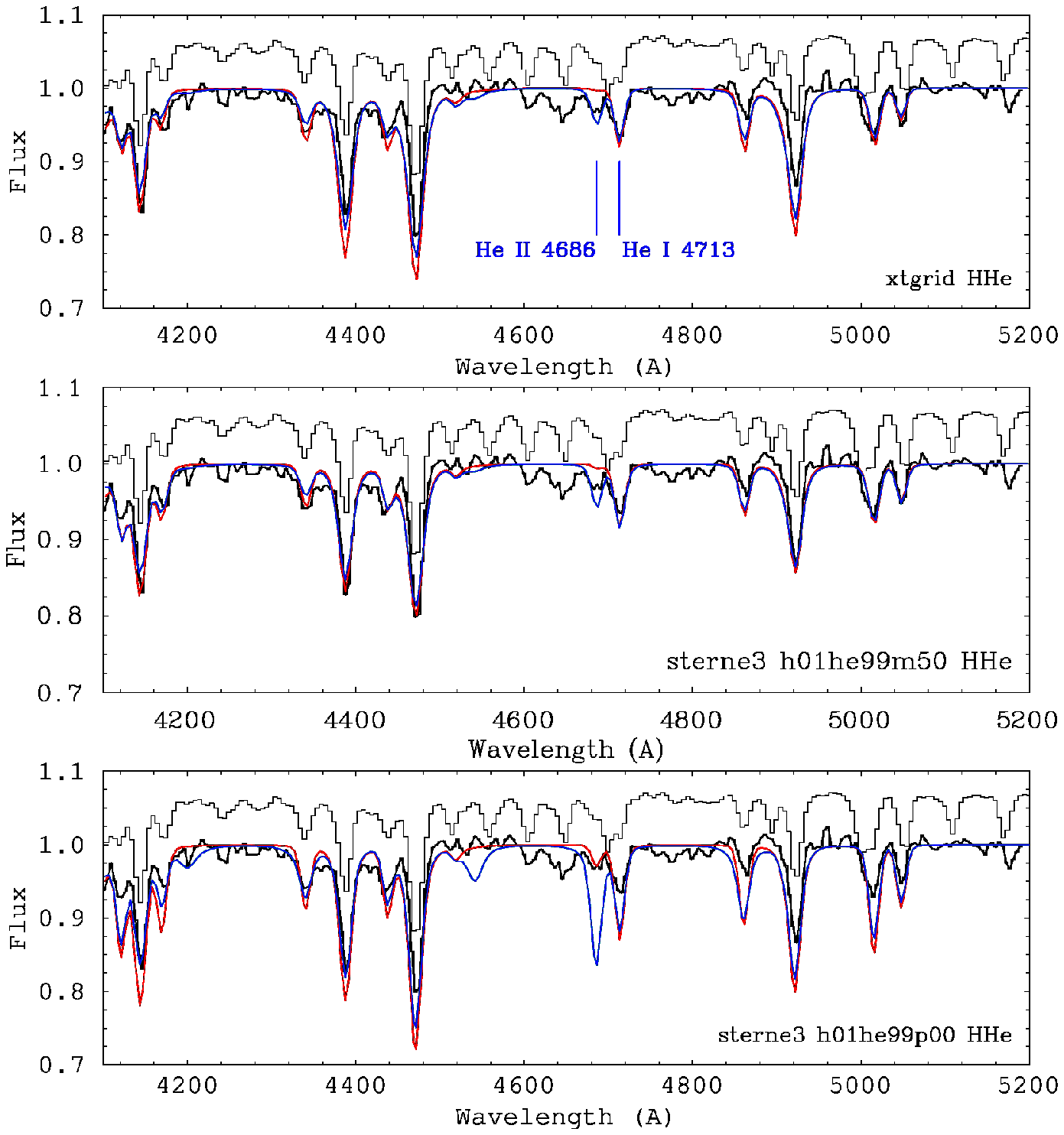,width=85mm}
        \caption{ Part of the EFOSC/NTT spectrum of \Jstar\ (bold black), normalised and compared with different model atmosphere predictions.    
In all three panels, the model spectra are for (\teff/kK,\lgcs,\yy) = (26.0,5.0,100) (red) and (36.0,5.5,100) (blue). The top panel shows model spectra  from a grid of non-LTE H+He  
models \citep{nemeth14}. The middle panel uses models computed using {\sc sterne3} with abundances of all elements other 
than H and He set to 0. The bottom panel uses models computed using {\sc sterne3} with abundances of all elements other 
than H and He set to solar values; the emergent spectrum includes H+He lines only.  The SALT-HRS spectrum is also shown, degraded to resolution 13\AA\  and offset vertically (light black). }
        \label{f:nemeth}
\end{figure}

\section[]{Analysis}
\label{s:fits}

A grid of model atmospheres and theoretical spectra was prepared based on the approximate composition for V652\,Her \citep{jeffery01b}.  
The input composition for the model grid was checked for consistency with the measured abundances, and an iteration was carried out where necessary. 
The observed spectrum used to fit  $T_{\rm eff}$  and  $g$  was the order-merged spectrum direct from the rectification procedure. 
This procedure was carefully designed to preserve the profiles of broad lines, even where they extend over substantial fractions of an order or across order overlaps. 
Initial estimates for $T_{\rm eff}$  and  $g$ were obtained using  the Armagh LTE\footnote{Local thermodynamic equilibrium was assumed throughout the analysis. Recent analyses of V652\,Her and the EHe BD$+10^{\circ}2179$ \citep{przybilla05,kupfer17} showed departures from LTE to be small except for very strong lines; the same is expected to be true here, and is evident in the final best-fit model.} model atmosphere codes {\sc sterne3}, {\sc spectrum} and {\sc sfit} \citep{behara06, jeffery17a}, by finding the best-fit spectrum in a grid which covers the solution space (\teff/kK,\lgcs,\yy) = (20(2)30, 2.8(0.2)4.4, 99).

The observed spectrum was further renormalised using the best-fit model so as to facilitate the measurement of abundances from narrow and weak lines (Figs.~\ref{f:lores}, \ref{f:atlasA} -- \ref{f:atlasC}). 
The renormalised spectrum was not used for measuring $T_{\rm eff}$ and $g$ since the wings of the crucial He\,{\sc i} lines are significantly affected by this renormalisation process. 

Inspection showed that  many lines were significantly sharper in the observed spectrum than in the model, with some close lines resolved in the former and not the latter. 
The rotation velocity was already set to zero. 
The instrumental profile was reduced to 0.001\AA\ (full-width half-maximum), corresponding to the nominal value for medium resolution observations with HRS ($37\,000$). 
In order to model the observed line splitting, the microturbulent velocity ($\xi$) in the model had to be reduced from  7 to 2 \kmsec. The value adopted for V652\,Her was 9 \kmsec, but dynamical effects in the pulsating atmosphere probably contribute to this value. 
Since the microturbulent velocity contributes significantly to the metal line opacity due and hence to the temperature structure of the model atmospheres, a new model grid was computed. 

\begin{figure*}
       \epsfig{file=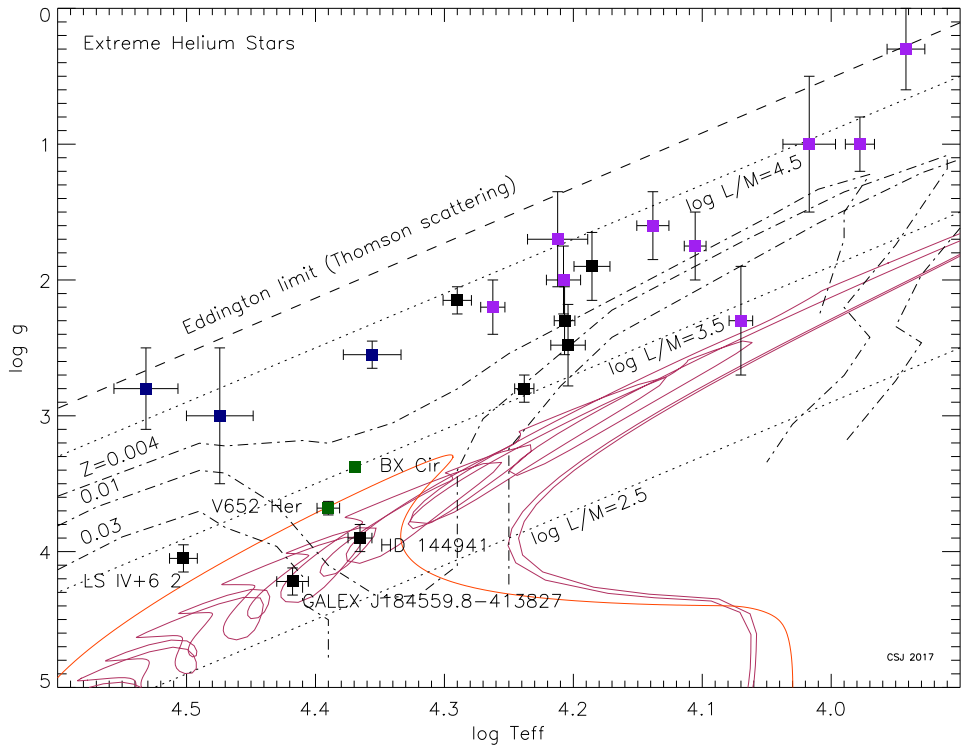,width=140mm}
        \caption{The $g-T_{\rm eff}$ diagram for all known extreme helium stars, including \Jstar. The
positions of the Eddington limit (Thomson scattering: dashed), 
luminosity-to-mass contours (solar units: dotted) and lower boundaries for pulsation 
instability (metallicities $Z$\,=\,0.004, 0.01, 0.03: dot-dashed) \citep{jeffery99} are also shown.
In the online version, variable EHes are shown in purple (cool), blue (hot), green (V652\,Her like variables). Non-variables are black. 
Data for $T_{\rm eff},g$ are as in \citet{jeffery08.ibvs}, except for BD+10$^{\circ}$2179 \citep{kupfer17}. 
Post-merger evolution tracks for models of 
He+He white dwarf mergers \citep[0.30+0.25 and 0.30+0.30\,$M_\odot$]{zhang12a} are shown in maroon.
Part of the post-flash track of a 0.46921\Msolar 'late hot flasher' (metallicity $Z=0.01$) is shown in orange \citep{bertolami08}.  
 } 
        \label{f:tracks}
\end{figure*}

The final surface properties measured for \Jstar\ are  $\teff = 26\,170 \pm750$\,K  and  $\lgcs= 4.22\pm 0.10$, with $\xi=2\kmsec$.
This is compared with other EHe stars in Fig.\,\ref{f:tracks}. 
An indication of the systematic error is obtained by the reduction in $\xi$, which produced shifts of 30\,K and 0.03 dex respectively.

Atmospheric abundances were obtained from the renormalised spectrum by $\chi^2$ minimisation, as in the analysis of UVO\,0825+15 \citep{jeffery17a}. 
Given the scatter seen in line-by-line analyses,  and the greater noise in the current spectrum, the abundance errors are conservatively estimated at $\pm 0.2$ dex for each species.   
The best-fit abundances of observed species are given in Table\,\ref{t:abunds}  as $\log \epsilon$, normalised to $\log \Sigma \mu \epsilon = 12.15$, 
where $\epsilon$ is the relative abundance by number of each species, and $\mu$ is its molecular weight. 
A partial atlas of the spectrum showing the best-fit model and marking the positions of all lines with equivalent widths in the model greater than 5\,m\AA\ is shown in Figs.\,\ref{f:atlasA} -- \ref{f:atlasC}.

\subsection{Previous work}

The above measurements of \teff\ and $g$ are sufficiently discrepant from those by \citet{vennes11} and \citet{nemeth12} to require comment
Although the EFOSC/NTT spectra is of very low resolution, it is sufficent to resolve the adjacent He{\sc i} 4713 \AA\ and He{\sc ii} 4686 \AA\
lines, which provide a critical temperature diagnostic at these temperatures. The online plot\footnote{http://stelweb.asu.cas.cz/$\sim$nemeth/work/galex/catalog/} of 
the \citet{nemeth12} fit to \Jstar\ shows both lines; although the fit is not perfect, the line ratio implies that the model at \teff = 36\,kK is consistent with the observation. 
The shape of the flux distribution is only a weak constraint since, for hot stars in the optical and near-ultraviolet, it is degenerate with the shape of interstellar extinction.  

The key to the discrepancy lies  in a large systematic difference between the degree of metal-line blanketing in the model atmospheres. 
\citet{vennes11} used non-LTE model atmospheres consisting of hydrogen and helium only. 
\citet{nemeth12} included carbon, nitrogen and oxygen, and obtained a very similar \teff.
The omission of opacity from  other ions, especially iron-group elements, has a major impact on the temperature structure of the model photosphere. 
This impact is further enhanced in hydrogen-deficient atmospheres owing to the absence of the normally dominant hydrogen photoionisation and the halving of the electron-scattering opacity (recall that $\kappa_{\rm es} = \kappa_{\rm es,0}/\mu_{\rm e} =  0.20(1+X)$, where $\mu_{\rm e}$ is the number of nucleons per free electron and $X$ is the hydrogen mass fraction). 
The {\sc sterne3} models are fully line blanketed and include opacity contributions from all major ions through to the iron-group and beyond \citep{behara06}.   

The consequence is illustrated in Fig.~\ref{f:nemeth}, where three model classes including (1) non-LTE H+He only, (2) LTE H+He only, and (3) LTE H+He+solar metals  are compared.
Two models are compared in each class, having \teff = 26 and 36 kK. The helium to hydrogen ratio is close to 100 in each case. 
The non-LTE models were taken from the grid of \citep{nemeth14}\footnote{http://www.ster.kuleuven.be/$\sim$petern/work/sd\_grid/}.
The LTE models were computed with {\sc sterne3} \citep{behara06}. 
To enable a proper comparison, the emergent spectra include H and He lines only. 
It is instructive to find (at this resolution) only minor differences between the non-LTE and LTE spectra where no metals are included in the model atmosphere calculation.
A very substantial shift in the He{\sc i}/He{\sc ii} ionization balance occurs when metals are included. 
This is due to the backwarming effect of metal-line opacity which heats the lower layers of the atmosphere and increases the degree of ionization in the line forming region at a given \teff. 
Hence  He{\sc ii} 4686 \AA\ appears at much lower \teff\ in fully-blanketed H-deficient models than it does in models which either have more hydrogen or lower metallicity. 
As demonstrated emphatically by \citet{anderson91} when discussing diagnostics for the atmospheres of normal B stars, it is essential to include all opacity sources before 
considering departures from LTE except, possibly, at the very lowest surface gravities.


\section{Discussion}

With $V=14.6$, \Jstar\ is 4 magnitudes fainter than V652\,Her. 
Its measured \Teff\ and $g$ imply $L/M$ approximately 3 times smaller.  
It is dangerous to infer a mass from putative evolutionary tracks for such stars. 
\citet{jeffery88} discussed core-mass shell-luminosity relations for helium-shell stars, but even  
assuming  a simple relation of the form $L \propto M^{\delta}$ is problematic since  $\delta$ varies strongly with mass. 
{\it Assuming} $M\approx 0.5\Msolar$, then  $L\approx  350 \Lsolar$ (with large errors). Also assuming extinction of 0.23 mag. for this position \citep{schlafly11}\footnote{http://irsa.ipac.caltech.edu/applications/DUST/} and a bolometric correction $-2.50$ mag., appropriate for early B stars, yields a distance $d\approx4.4 $\,kpc. With galactic coordinates  $l^{\sc ii}=354.1900, b^{\sc ii}= -16.6900$, this gives a location half way toward the Galactic bulge, with height $z \approx 1.4 $\,kpc below the plane. 

The spectroscopic similarities with V652\,Her argue for a similar evolutionary status; in that case the  model most consistent with observations (including pulsations) is that of a post double helium white dwarf merger \citep{saio00,zhang12a}. 
Tracks for  models of such stars are shown in Fig.~\ref{f:tracks}. If the same is true here, \Jstar\ will evolve to become a helium and nitrogen-rich  hot subdwarf  within 10$^5$\,y or so. 
The enrichment of nitrogen and depletion of carbon point to a highly CNO-processed helium surface; some hydrogen  from surface layers of the progenitor white dwarfs has survived \citep{hall16}. 
The photospheric iron abundance is subsolar ($\approx 0.4$ dex), but not low enough to be considered `metal-poor'.  
 The merger of a CO white dwarf and a helium white dwarf is predicted to produce a carbon-rich surface, having negligible hydrogen (typically $<0.01\%$ by number), and a much higher $L/M$ ratio than observed in the current case \citep{saio02}. 

A second class of model which might apply to such stars includes the `late hot flasher' models wherein helium-core ignition occurs some time after a star leaves the red giant branch \citep{brown01,bertolami08}; if the remaining hydrogen envelope is sufficiently thin, flash-driven convection can enrich the surface with CNO-processed helium as well as carbon. 
The evolution track for a  $Z=0.01$, 0.46921\Msolar\ model is shown in Fig.~\ref{f:tracks}. 
Neither model satisfactorily accounts for the mass loss required for the star to leave the giant branch before helium ignition, unless assisted by a binary or planetary companion. 
\citet{brown01} predicted high carbon abundances. 
\citet{bertolami08} found a range of surface mixtures from (for example)  high carbon (4\% by mass, initial metallicity $Z=0.01$), negligible nitrogen at high mass (0.47725\Msolar), to  moderate carbon (1.2\%) and nitrogen  (2.1\%) at lower mass (0.46644\Msolar), for models with deep mixing, of which the  0.47725\Msolar\ model passes closest to the position of \Jstar.
There is no evidence  that this class of models can produce pre-subdwarfs with CNO-cycled helium surfaces and without carbon enrichment.

There is a clear distinction between the nitrogen-rich, carbon-poor surface of \Jstar\ and the carbon-rich surfaces of all EHe's other than V652\,Her \citep[][Table 1]{jeffery11a}, 
where the mean carbon abundance is $\approx2.5$ dex higher than in \Jstar.  
The same is true of all RCrB stars, to which high luminosity EHe's are thought to be related, though the difference here is  $\approx2$ dex (ibid.). 
Three other relatively hot `high-gravity' helium stars are shown in Fig.\,\ref{f:tracks} and Table\,\ref{t:abunds}. 
LS\,IV$+6^{\circ}2$ and BX\,Cir are both carbon-rich and hydrogen-poor relative to V652\,Her and \Jstar. 
HD144941 is more hydrogen-rich and extremely metal poor. It is not  clear how or whether these stars are related to one another. 
Amongst  helium-rich hot subdwarfs, there exists a range of carbon abundance from  $\log \epsilon_{\rm C} = 6.73\pm0.18$ (SB\,21) to 8.94 (BPS CS 29496--0010), 
whilst the nitrogen abundances are clustered around $\log \epsilon_{\rm N} \approx 8.4$ \citep{hirsch09,naslim10}. 

The question of whether \Jstar\ also pulsates has yet to be determined;  \Jstar\ lies outside the  region where radial pulsations driven by the Z-bump opacity mechanism are predicted \citep{jeffery99a}.
Due to its smaller radius, any radial or p-mode pulsation period must be shorter than that of V652\,Her (0.108\,d). 
There is no evidence of any  periodic variability in the All Sky Automated Survey (ASAS) catalogue entry for this object \citep{pojmanski05}. 
Since the ASAS cadence is $<1 {\rm d}^{-1}$, this is not a strong non-detection. 

A remarkable property of \Jstar\ is the sharpness of its absorption lines. 
This places strong constraints on both the rotation velocity and the microturbulent velocity in the photosphere. 
One question posed by the hypothesis for its origin in a  white dwarf merger origin is how \Jstar\ could have lost so much angular momentum during the immediate post-merger evolution. 
On the other hand, a significant number of helium-enriched subdwarf B stars are being identified with very low rotation velocities \citep{naslim11,naslim13,jeffery17a}.  

\section{Conclusion}

SALT HRS observations of a faint-blue star previously classified He-sdB \citep{vennes11} demonstrate
it to be a nitrogen-rich EHe star similar to the pulsating EHe star V652\,Her. 
\Jstar\ becomes the first EHe star to be discovered for nearly 40 years, indicative of the extreme rarity of these stars.  
Its surface is predominantly that of CNO-processed helium with some hydrogen contamination, pointing to a possible origin in a double helium white dwarf merger. 
The Galactic position and metallicity of \Jstar\ are compatible with membership of the bulge population. 
This discovery suggests that there are more EHe stars waiting to be found, at least with relatively low luminosities.
A higher signal-to-noise spectrum will allow abundances and other parameters to be refined further. 

\section*{Acknowledgments}

The observation reported in this paper was obtained with the Southern African Large Telescope (SALT)
under program 2016-2-SCI-008 (PI: Jeffery). The author is indebted to the hard work of the entire SALT team.
He thanks Peter N\'emeth for a copy of the reduced EFOSC/NTT spectrum of \Jstar. 

The Armagh Observatory is funded by direct grant form the Northern Ireland Dept of Communities.
CSJ acknowledges support from the UK Science and Technology Facilities Council (STFC) Grant No. 
ST/M000834/1. 

This research has made use of the SIMBAD database, operated at CDS, Strasbourg, France.

\bibliographystyle{mn2e}
\bibliography{ehe}

\appendix
\renewcommand\thefigure{A.\arabic{figure}} 
\renewcommand\thetable{A.\arabic{table}} 

\section[]{Spectral Atlas  for \Jstar}
\label{s:app1}
\label{s:lines}
\begin{figure*}
\epsfig{file=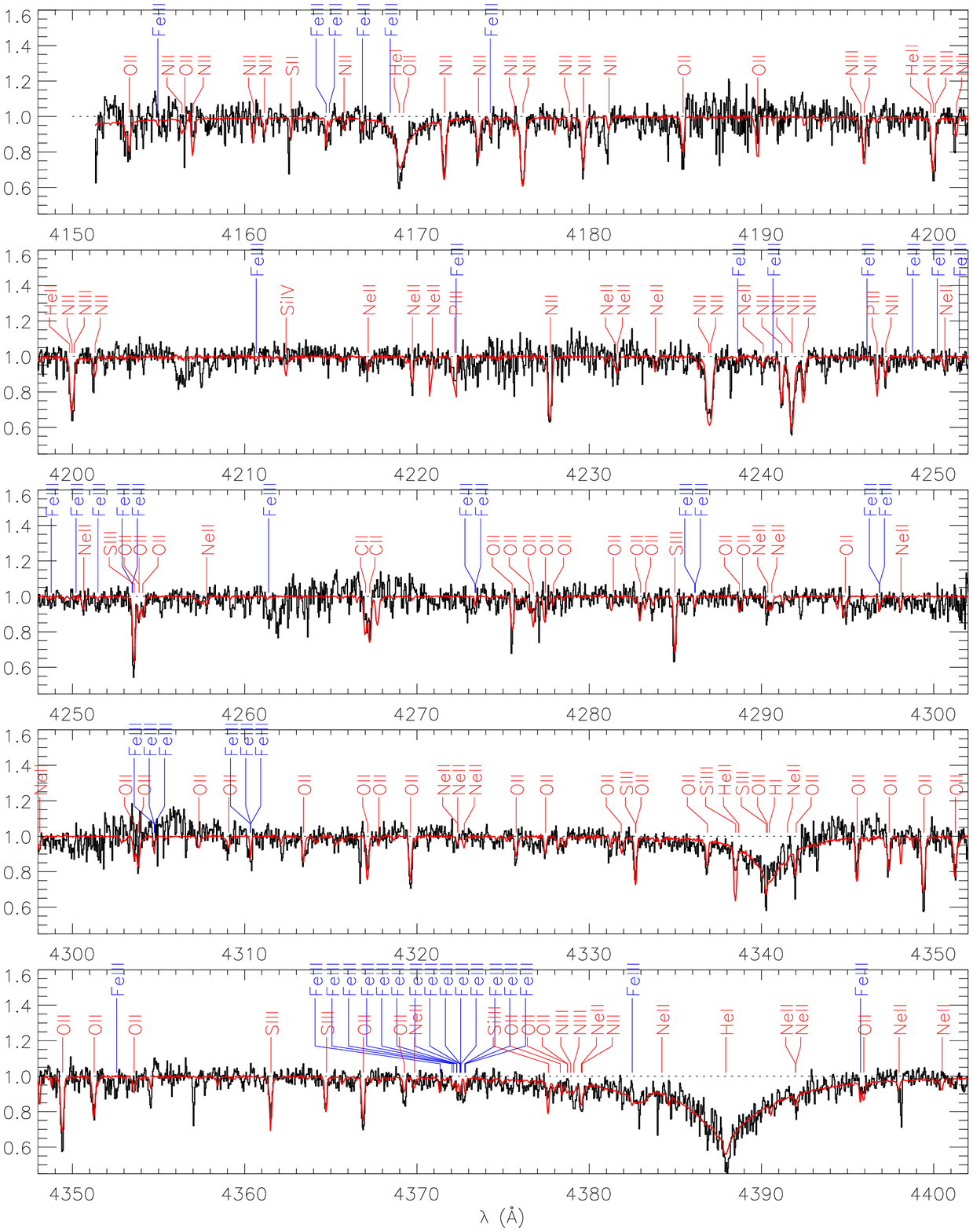,width=0.9\textwidth}
\caption{Parts of the observed SALT HRS spectrum of \Jstar\ (black histogram), 
and the best-fit model having  $\teff=26\,000$\,K, $\lgcs=4.2$),  $n_{\rm H}/n_{\rm He}=0.01$ 
and abundances shown in Table~\ref{t:abunds} (red polyline).
Lines with theoretical equivalent widths greater than $5$m\AA\ are identified wherever possible. 
Apparently large bins in the observed spectrum correspond to major instrumental artefacts; other artefacts appear as very sharp regularly-spaced lines.}
\label{f:atlasA}
\end{figure*}

\begin{figure*}
\epsfig{file=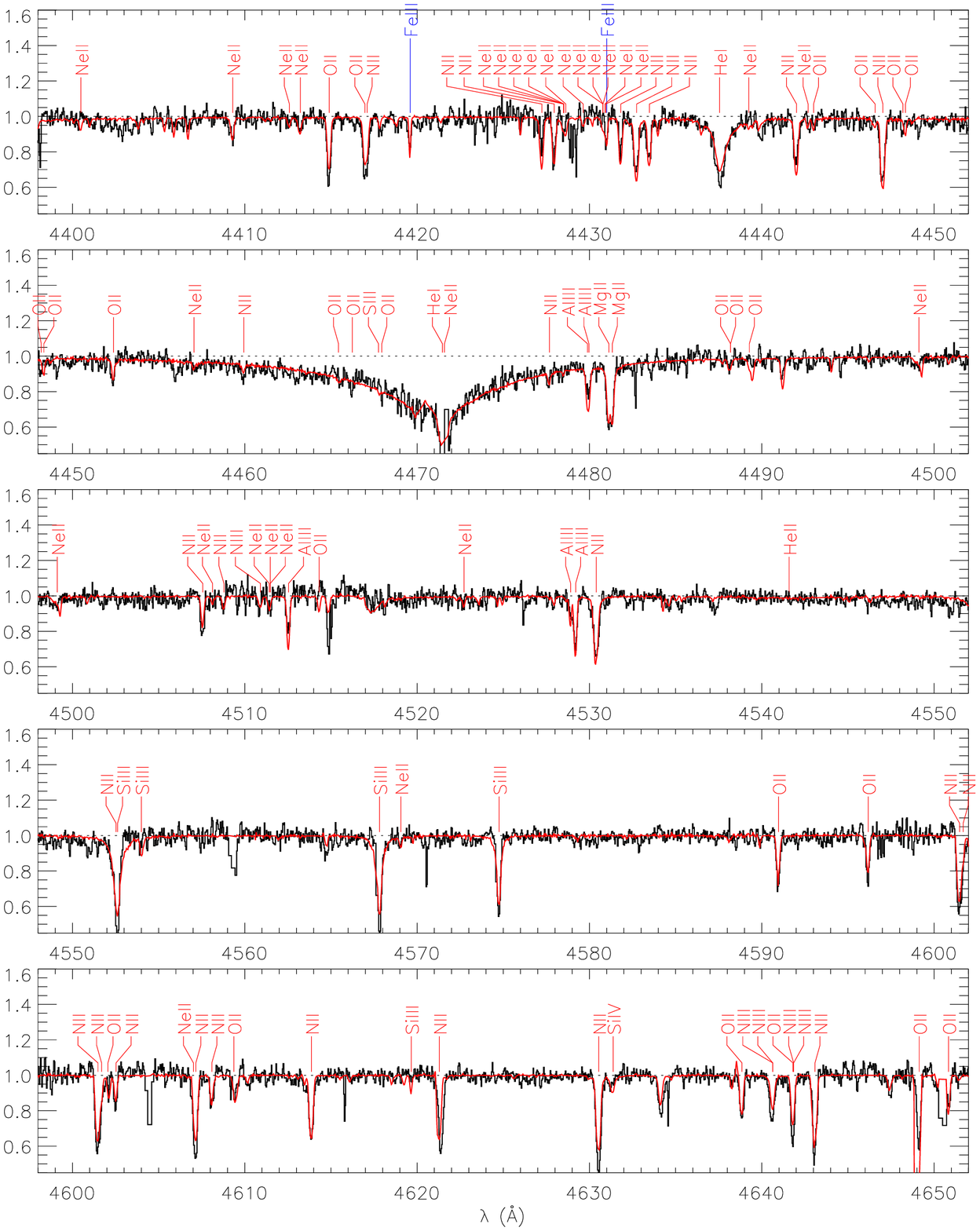,width=0.9\textwidth}
\caption{As Fig.~\ref{f:atlasA} (contd.)}
\label{f:atlasB}
\end{figure*}

\begin{figure*}
\epsfig{file=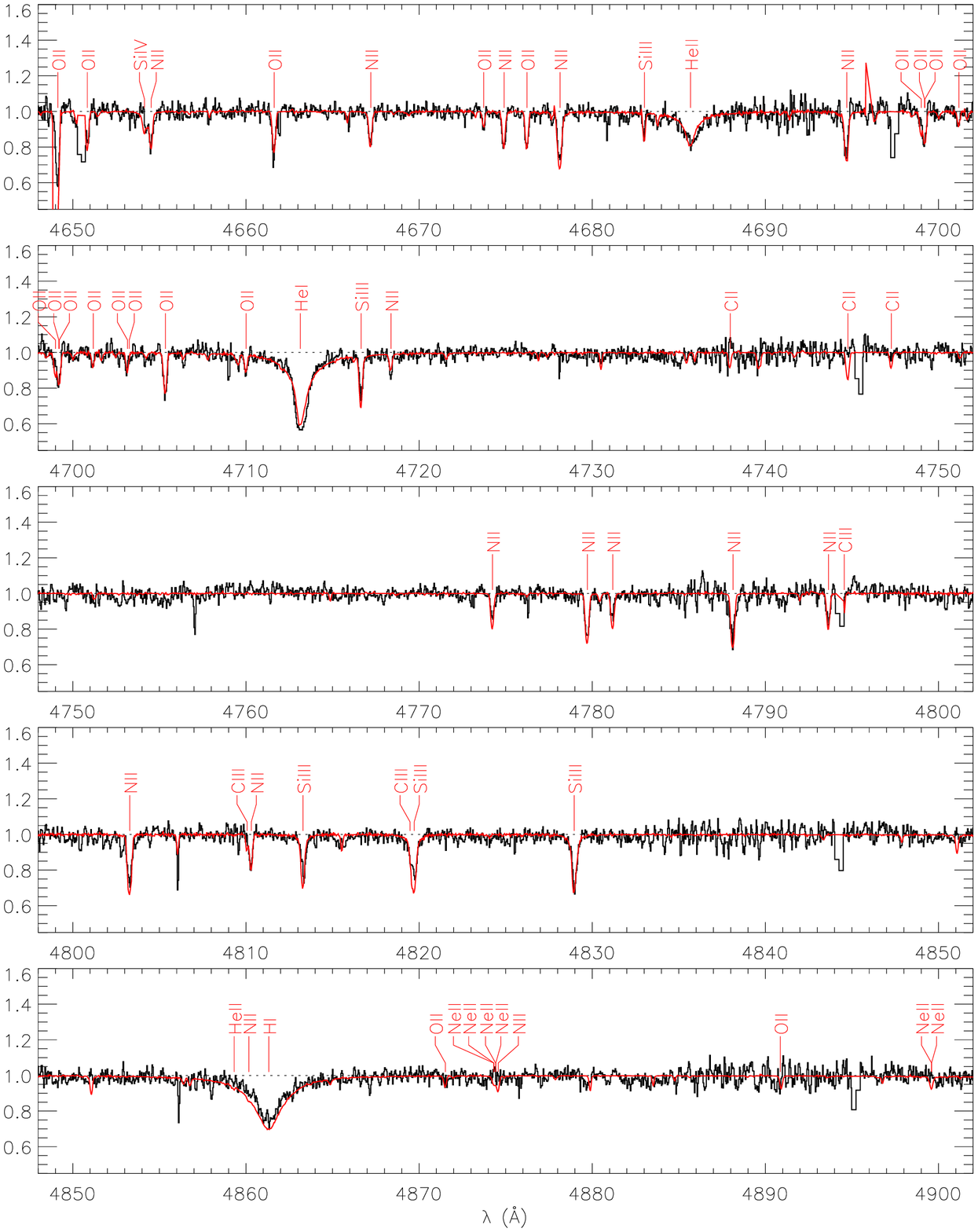,width=0.9\textwidth}
\caption{As Fig.~\ref{f:atlasA} (contd.)}
\label{f:atlasC}
\end{figure*}

Figures \ref{f:atlasA} to \ref{f:atlasC} contain a partial atlas of the {\it SALT HRS} spectrum of \Jstar\
with the best model fit and identifications of absorption lines.

%

\label{lastpage}
\end{document}